\documentclass[12pt]{article}
\usepackage{graphicx}
\usepackage{latexsym}
\usepackage{amsmath}
\usepackage{amssymb}
\usepackage{bm}
\usepackage[authoryear]{natbib}

\renewcommand{\appendix}{\par
        \setcounter{section}{0}
        \def\thesection{Appendix.}
        \def\theequation{\arabic{equation}}
}

\title{Short-term Synaptic Depression Improves Error-correcting Ability in Cortical Circuits}
\author{Narihisa Matsumoto$^{1}$, Daisuke Ide$^{2}$, \\
Masataka Watanabe$^{2}$ and Masato Okada$^{3,4}$}

\date{
$^{1}$Neuroscience Research Institute, National Institute of Advanced Industrial Science and Technology (AIST), \\Ibaraki 305-8568, Japan\\
$^{2}$Faculty of Engineering, University of Tokyo, \\Tokyo 113-0033,Japan\\
$^{3}$Graduate School of Frontier Sciences, University of Tokyo, \\Chiba 277-8561, Japan\\
$^{4}$PRESTO, Japan Science and Technology Agency, \\Chiba 277-8561, Japan\\
}

\begin{document}
\maketitle
\newpage
\section*{abstract}
Synaptic connections are known to change dynamically. 
High-frequency presynaptic inputs induce decrease of synaptic weights. 
This process is known as short-term synaptic depression. 
The synaptic depression controls a gain for presynaptic inputs.
However, it remains a controversial issue what are functional roles of this gain control.
We propose a new hypothesis that one of the functional roles is to enlarge basins of attraction. 
To verify this hypothesis, we employ a binary discrete-time associative memory model which consists of excitatory and inhibitory neurons. 
It is known that the excitatory-inhibitory balance controls an overall activity of the network.
The synaptic depression might incorporate an activity control mechanism.
Using a mean-field theory and computer simulations, we find that the basins of attraction are enlarged whereas the storage capacity does not change.
Furthermore, the excitatory-inhibitory balance and the synaptic depression work cooperatively.
This result suggests that the synaptic depression works to improve an error-correcting ability in cortical circuits.

\newpage
\section{Introduction}
Cortical neurons receive thousands of synaptic inputs.
The neurons might receive high-frequency presynaptic inputs.
Extracting intrinsic signals from random fluctuations in high-frequency presynaptic inputs is a severe challenge.
Some mechanisms that reduce a gain for presynaptic inputs might work at synaptic sites.
Such gain control cannot be achieved through fixed synaptic weights because input rates change dynamically.
Neurophysiological experiments show that high-frequency inputs induce the decrease of synaptic weights \cite[]{Thomson94}.
This process is known as short-term synaptic depression.
The synaptic depression is known to control the gain for presynaptic inputs \cite[]{Abbott97,Tsodyks97}.
This property might influence not only the activity of a single neuron but also that of the overall activity \cite[]{Bressloff99,Kistler99,Tsodyks00}.

However, it is still a controversial issue what are functional roles of the gain control.
To elucidate the functional roles, some information must be embedded in the synaptic connections.
We employ an associative memory model that stores memory patterns in synaptic connections.
Only a few works have investigated how the synaptic depression affects the performance of the associative memory model \cite[]{Bibitchkov02,Pantic02,Torres02}.
The memory patterns embedded by Hebb rule \cite[]{Hebb49} become fixed points, i.e., attractors \cite[]{Amit89}.
Bibitchkov et al. found that the synaptic depression reduced a storage capacity \cite[]{Bibitchkov02}.
Torres et al. found that the storage capacity decreased with the degree of the depression in a thermodynamical limit \cite[]{Torres02}.
However, the main targets of these works were the steady states of the models.
It is necessary to investigate the dynamical properties of the model because the synapses change dynamically.
One of the important dynamical properties is basins of attraction which express the regions where the system converges to stored patterns.
In the view of information processing, the basins of attraction reflect an error-correcting ability.
Bibitchkov et al. found that the synaptic depression \textit{shrunk} the basins of attraction at a large loading rate and increased them a little at a small loading rate \cite[]{Bibitchkov02}.
In contrast, we propose a new hypothesis that the synaptic depression \textit{enlarges} the basins of attraction not only at a small loading rate but at a large loading rate.

We employ a binary discrete-time associative memory model which consists of excitatory and inhibitory neurons.
The memory patterns with a small firing rate, i.e., sparse patterns, are embedded in the synaptic connections among the excitatory neurons.
This coding scheme is known as sparse coding.
An activity control is requisite for the stable retrieval of the sparse patterns \cite[]{Okada96}.
The excitatory-inhibitory balance controls an overall activity.
The synaptic depression might incorporate the activity control mechanism.
Specifically, when an overall activity is high, the synaptic depression decreases the gain for presynaptic inputs and maintains the overall activity at a constant level.
We also investigate whether the excitatory-inhibitory balance and the synaptic depression work cooperatively or not.

This paper consists of seven sections.
The section 2 gives the description of the model employed in this paper.
In the section 3 mean-field equations which describe a steady state of the model are shown.
In the sections 4 and 5 we investigate how the synaptic depression influences the performance of the model which consists of only excitatory neurons.
In the section 6 we investigate the relationship between the synaptic depression and the excitatory-inhibitory balance.
In the section 7 we summarize the results obtained in this paper and make a discussion about the model.

\section{Model}
The model used in this study consists of excitatory and inhibitory neurons \cite[]{Amit94,Vreeswijk96,Matsumoto05}.
The $i$-th excitatory neuron ($i=1,\cdots,N$) is characterized by its binary state $s_i(t)=\{0,1\}$ and discrete time $t$.
If the excitatory neuron fires at time $t$, its state is $s_i(t)=1$; otherwise, $s_i(t)=0$.
A thermodynamics limit, $N \rightarrow \infty$, is considered.
The excitatory neurons are all-to-all connected.
The synaptic weight from the presynaptic excitatory neuron $j$ to the postsynaptic excitatory neuron $i$, $J_{ij}(t)$, are dynamically changed, and its specific value will be discussed later.
The synaptic weights between the excitatory neurons and the inhibitory neurons are uniform.
There are no connections among the inhibitory neurons.
Therefore, the population of the inhibitory neurons can be regarded as a single inhibitory neuron.
The state of the $i$-th excitatory neuron is updated by the synchronous rule:
\begin{eqnarray}
s_i(t+1) &=& \Theta\Big( h_i(t) - g\big(\bar{s}(t) -f \big) - \hat{\theta} \Big) \label{eq.model} \\
&=& \Theta\Big( \sum^N_{j \ne i} J_{ij}(t)s_j(t) - g\big(\bar{s}(t) -f \big) - \hat{\theta} \Big) \\
&=& \Theta\Big( \sum^N_{j \ne i} J_{ij}(t)s_j(t) - g I(t) - \hat{\theta} \Big) \\
&=& \Theta\Big( \sum^N_{j \ne i} J_{ij}(t)s_j(t) - \bar{\theta}(t) \Big) 
\end{eqnarray}
where $h_i(t)=\sum^N_{j \ne i} J_{ij}(t)s_j(t)$ denotes an internal potential, $\bar{s}(t)=\frac{1}{N}\sum^N_{j=1} s_j(t)$ denotes a mean firing rate of the excitatory neurons, $f$ denotes a firing rate of memory patterns, $g$ denotes the strength of the inhibition, $\hat{\theta}$ is a uniform threshold, and $\bar{\theta}(t)=g I(t) + \hat{\theta}$ is an effective threshold.
The inhibitory neuron receives the mean output of the excitatory neurons, $\bar{s}(t)$, and it sends output to the excitatory neurons as $I(t)=\bar{s}(t)-f$.
$g=0$ means that the model consists of only excitatory neurons.
If the mean firing rate of the excitatory neurons, $\bar{s}(t)$, is higher than $f$, the inhibition increases an effective threshold $\bar{\theta}(t)$.
Then the excitatory neurons tend to be silent.
On the contrary, if $\bar{s}(t)$ is lower than $f$, the inhibition decreases the effective threshold $\bar{\theta}(t)$.
Then the excitatory neurons tend to activate.
The output function $\Theta(\cdot)$ of the excitatory neurons is a step function:
\begin{eqnarray}
\Theta(u) &=&
\begin{cases}
1.   &(u \ge 0)\\
0.   &(u<0)
\end{cases}
\end{eqnarray}
Memory patterns $\bm{\xi}^{\mu}$ ($\mu=1,2,\cdots,p$) are stored in the synaptic connections among the excitatory neurons.
Each element $\xi_i^{\mu}$ of the memory pattern $\bm{\xi}^{\mu}=(\xi^{\mu}_1,\xi^{\mu}_2,\cdots,\xi^{\mu}_N)$ is generated independently by
\begin{equation}
\mbox{Prob}[\xi_i^{\mu}= 1]=1-\mbox{Prob}[\xi_i^{\mu}= 0]=f.
\end{equation}
The expectation of $\bm{\xi}^{\mu}$ is $\mbox{E}[\xi_i^{\mu}]=f$.
We consider a small firing rate $f$.
A value $\alpha=p/N$ is defined as a loading rate.
When the loading rate $\alpha$ is larger than a critical value $\alpha_C$, the retrieval of memory patterns become unstable.
The critical value $\alpha_C$ is known as a storage capacity.

The closeness between $\bm{s}(t)$ and $\bm{\xi}^{\mu}$ at time $t$ is characterized by an overlap
\begin{equation}
m^{\mu}(t)=\frac{1}{Nf(1-f)}\sum_{i=1}^N(\xi^{\mu}_i-f)s_i(t).
\end{equation}
If the overlap is close to $1$, i.e., $m^{\mu}(t) \approx 1$, the model succeeds to retrieve the memory pattern $\bm{\xi}^{\mu}$.
Hereafter, the target pattern for the retrieval is the first memory pattern $\bm{\xi}^1$.

The synaptic weight $J_{ij}(t)$ incorporating the synaptic depression is determined by a phenomenological model of synapses \cite[]{Abbott97,Tsodyks97}.
When the synapses transmit input signals, they exhaust a finite amount of resources, e.g., neuromodulators.
A dynamical amplitude factor $x_j(t)$ denotes the fraction of available resources.
After each spike is emitted, the resources decrease by a certain fraction $U_{SE}$ and to recover with a time constant $\tau$.
If the recover lags behind the interval of high-frequency presynaptic inputs, the amount of resources decreases and the synapses are depressed.
The factor $x_j(t)$ is updated by the synaptic dynamics \cite[]{Tsodyks97,Pantic02}:
\begin{equation}
x_j(t+1)=x_j(t)+\frac{1-x_j(t)}{\tau}-U_{SE}x_j(t)s_j(t). \label{eq.depression}
\end{equation}
$x_j(t)=1$ implies that the synapses are not depressed.
The synaptic weight $J_{ij}(t)$ incorporating the synaptic depression is obtained by multiplying a fixed synaptic weight $\tilde{J}_{ij}$ and the dynamic amplitude factor $x_j(t)$ ($0<x_j(t)\leq1$):
\begin{equation}
J_{ij}(t)=\tilde{J}_{ij}x_j(t).
\end{equation}
The synaptic weight $\tilde{J}_{ij}$ is determined by a Hebbian-like rule, i.e., a covariance rule:
\begin{equation}
\tilde{J}_{ij} = \frac{1}{Nf(1-f)}\sum^p_{\mu =1}(\xi^{\mu}_i-f)(\xi^{\mu}_j-f).
\end{equation}
A self-connection $\tilde{J}_{ii}$ is assumed to be nonexistent.
For simplicity, the synaptic depression is incorporated into only excitatory-excitatory connections.
The excitatory-inhibitory connections are fixed.

\section{Mean-field Equations at a Steady State}
In this section, we derive mean-field equations at a steady state, i.e., $t \rightarrow \infty$.
For simplicity, individual values at the steady state are written by $x_j(\infty)=x_j$, $h_i(\infty)=h_i$, $s_i(\infty)=s_i$, and $m^{\mu}(\infty) = m^{\mu}$.
The factor $x_j(t)$ reaches its steady-state value by $t \rightarrow \infty$ in equation (\ref{eq.depression}) \cite[]{Bibitchkov02}:
\begin{equation}
x_j=\frac{1}{1+\gamma s_j},
\end{equation}
where $\gamma=\tau U_{SE}$.
The value $\gamma$ indicates the level of the synaptic depression.
Since $s_j$ takes a binary value, i.e., $s_j = \{0,1\}$, the value $x_j s_j$ can be written as
\begin{equation}
x_j s_j=\frac{s_j}{1+\gamma s_j}=\frac{1}{1+\gamma}s_j.
\end{equation}
By using this relationship, the internal potential $h_i$ at the steady state is written as
\begin{equation}
h_i = \sum_{j \ne i}^N \tilde{J}_{ij}x_j s_j = \sum_{j \ne i}^N \tilde{J}_{ij}\frac{s_j}{1+\gamma} = \frac{1}{1+\gamma}\sum_{j \ne i}^N \tilde{J}_{ij}s_j.
\end{equation}
Then, the neuronal state $s_i$ is written as
\begin{equation}
s_i = \Theta( h_i - g(\bar{s} -f ) - \hat{\theta} ) = \Theta\Big( \frac{1}{1+\gamma}\sum_{j \ne i}^N \tilde{J}_{ij}s_j - g(\bar{s} -f ) - \hat{\theta}\Big).
\end{equation}
By using equations (7), (9), and (12), the internal potential $h_i$ is represented as
\begin{eqnarray}
h_i \!\!\!\!&=&\!\!\!\! \frac{1}{1+\gamma}\sum_{j \ne i}^N\tilde{J}_{ij}s_j 
= \frac{1}{1+\gamma}\frac{1}{Nf(1-f)}\sum^p_{\mu =1}\sum^N_{j \ne i}(\xi^{\mu}_i-f)(\xi^{\mu}_j-f)s_j\\
\!\!\!\!&=&\!\!\!\! \frac{1}{1+\gamma}\bigg\{ \frac{1}{Nf(1-f)}\sum^N_{j=1}(\xi^1_i-f)(\xi^1_j-f)s_j\nonumber \\
\!\!\!\!& &\!\!\!\! +\frac{1}{Nf(1-f)}\sum^p_{\mu =2}\sum^N_{j=1}(\xi^{\mu}_i-f)(\xi^{\mu}_j-f)s_j
                    -\frac{1}{Nf(1-f)} \sum^p_{\mu=1} (\xi^{\mu}_i-f)^2s_i \bigg\} \nonumber\\ \\
\!\!\!\!&=&\!\!\!\! \frac{1}{1+\gamma} \bigg\{ (\xi^1_i-f)m^1+z_i\bigg\},\\
z_i \!\!\!\!&=&\!\!\!\! \sum_{\mu=2}^p(\xi^{\mu}_i-f)m^{\mu} -\alpha s_i.
\end{eqnarray}
The first term of the equation (17) is a signal term for the retrieval of the target pattern $\bm{\xi}^1$.
The second term is a cross-talk noise term which represents contributions from non-target patterns and prevents the target pattern $\bm{\xi}^1$ from being retrieved.
According to a mean-field theory \cite[]{Okada96,Shiino92}, the cross-talk noise obeys a Gaussian distribution with mean $\Gamma s_i$ and variance $\sigma^2$.
By using this theory, the neuronal state is written as
\begin{equation}
s_i = \Theta \Big(\frac{1}{1+\gamma}\left((\xi^1-f)m^1 + \sigma z_i + \Gamma s_i \right) - g(\bar{s} - f) - \hat{\theta} \Big).
\end{equation}
By applying Maxwell rule to equation (19), the solution $s_i$ is obtained as
\begin{equation}
s_i = \Theta \Big(\frac{1}{1+\gamma}\left((\xi^1-f)m^1 + \sigma z_i + \frac{\Gamma}{2} \right) - g(\bar{s} - f) - \hat{\theta} \Big).
\end{equation}
The mean-field equations describing the steady state of the model are obtained by the following equations.
For simplicity, the overlap $m^1$ is written as $m$.
\begin{eqnarray}
m &=& -\frac{1}{2} \mathrm{erf}(\phi_1) + \frac{1}{2} \mathrm{erf}(\phi_2),\\
U &=& \frac{f}{\sqrt{2\pi} \sigma}\exp(-\phi_1^2) + \frac{1-f}{\sqrt{2\pi} \sigma}\exp(-\phi_2^2),\\
q &=& \frac{1}{2}-\frac{f}{2}\mathrm{erf}(\phi_1) - \frac{1-f}{2}\mathrm{erf}(\phi_2),\\
\bar{s} &=& \frac{1}{2}-\frac{f}{2}\mathrm{erf}(\phi_1) - \frac{1-f}{2}\mathrm{erf}(\phi_2),
\end{eqnarray}
where $\phi_1=\frac{-(1-f)m+(1+\gamma)g(\bar{s}-f)+(1+\gamma)\hat{\theta}-\frac{\Gamma}{2}}{\sqrt{2 \alpha \sigma^2}}$, $\phi_2=\frac{fm+(1+\gamma)g(\bar{s}-f)+(1+\gamma)\hat{\theta}-\frac{\Gamma}{2}}{\sqrt{2 \alpha \sigma^2}}$, $\sigma^2 = \frac{\alpha q}{(1-U)^2}$, $\Gamma=\frac{\alpha U}{1-U}$, $\mathrm{erf}(y) = \frac{2}{\sqrt{\pi}}\int_0^y\mathrm{exp}(-u^2)du$.
Solving these equations numerically, we discuss the macroscopic state of the model.
The detailed derivation of the mean-field equations is shown in appendix.

The mean-field equations derived in this paper are different from the equations in the previous works \cite[]{Bibitchkov02,Torres02}.
The equations that Bibitchkov et al. derived dropped out the equation of $U$ (equation (22) in this paper).
Therefore, the cross-talk noise was not estimated accurately.
Torres et al. assumed that $x_j$ was independent of $s_j$.
However, the equation (\ref{eq.depression}) apparently shows that $x_j$ depends on $s_j$.
Therefore, this assumption was invalid.
Thus, the mean-field equations derived in this paper can describe the steady state of the model more accurately than in the previous works.

\section{Storage Capacity}
We investigate how the synaptic depression influences the steady state of the model.
In this section, we consider excitatory neurons, i.e., $g=0$.
Since the output function in equation (\ref{eq.model}) is a step function, the neuronal state is determined by the sign of an argument.
Then, the neuronal state at the steady state is written as
\begin{align}
s_i &= \Theta\Big( \frac{1}{1+\gamma}\sum_{j \ne i}^N \tilde{J}_{ij}s_j -\hat{\theta}\Big)
= \Theta\Big(\frac{1}{1+\gamma}\big(\sum_{j \ne i}^N \tilde{J}_{ij}s_j -(1+\gamma)\hat{\theta}\big)\Big)\\
&= \Theta\Big( \sum_{j \ne i}^N \tilde{J}_{ij}s_j -(1+\gamma)\hat{\theta}\Big).
\end{align}
If the threshold $\hat{\theta}$ is set at $\hat{\theta}=\theta/(1+\gamma)$, the neuronal state at the steady state is written as
\begin{equation}
s_i = \Theta\Big( \sum_{j \ne i}^N \tilde{J}_{ij}s_j -\theta\Big),
\end{equation}
where $\theta$ is a threshold when the synaptic depression is not incorporated into the model.
This equation implies that the steady state of the model incorporating the synaptic depression is equivalent to the one without the synaptic depression when the threshold is set at $\hat{\theta}=\theta/(1+\gamma)$.
Hereafter, the threshold with the synaptic depression is written as $\hat{\theta}$ while the threshold without the synaptic depression as $\theta$.

Here, we show the mechanisms where the memory pattern $\bm{\xi}^1$ is retrieved in the model with the synaptic depression.
From the equation (20), the internal potential at the steady state is given by
\begin{equation}
h_i = \frac{1}{1+\gamma}\left\{(\xi^1-f)m^1 + \sigma z_i + \frac{\Gamma}{2} \right\}.
\end{equation}
At first, we consider the case where the number of memory patterns is small, i.e., $p \sim O(1)$ and the synapses are not depressed, i.e., $\gamma = 0$.
The cross-talk noise does not exist in this case, and the second and third terms in the equation (28) vanish because these terms come from the cross-talk noise.
Let the neuronal state at the steady state be equivalent to the first memory pattern: $\bm{s}=\bm{\xi}^1$.
The probability distribution of $h_i$ is shown in Figure \ref{fig1}(a).
When the threshold $\theta$ is set between $-f$ and $1-f$, the neuronal state $s_i$ takes $1$ with probability $f$ and $0$ with $1-f$.
The retrieval of the memory pattern $\bm{\xi}^1$ is successful.

Next, we consider the case where the number of memory patterns is order $N$, i.e., $p \sim O(N)$ and the synapses are not depressed, i.e., $\gamma = 0$.
The probability distribution of $h_i$ is shown in Figure \ref{fig1}(b).
Setting the threshold $\theta$ at an appropriate value is essential for the stable retrieval \cite[]{Okada96,Matsumoto02}.
Finally, we consider the case where the number of memory patterns is order $N$, i.e., $p \sim O(N)$ and the synapses are depressed, i.e., $\gamma>0$.
Since the signal and the variance of the cross-talk noise are scaled by $1/(1+\gamma)$ (see equation (28)), the probability distribution of $h_i$ is shown in Figure \ref{fig1}(c).
When the threshold $\hat{\theta}$ is set at $\hat{\theta}=\theta/(1+\gamma)$, the retrieval of the memory pattern $\bm{\xi}^1$ succeeds.

Solving the mean-field equations (equations (21-24)) numerically, the steady state of the model can be analyzed.
Figure \ref{fig2}(a) shows the dependency of the overlap $m$ on the loading rate $\alpha$ without the synaptic depression, i.e., $\gamma=0$ at $f=0.1$.
The dashed lines are obtained by solving the mean-field equations (equations (21-24)) numerically while the error bars indicate medians and quartile deviations of $m(100)$ obtained by computer simulations in $11$ trials at $N=5000$.
The initial state is set at the first memory pattern, i.e., $\bm{s}(0)=\bm{\xi}^1$, and the threshold is fixed at $\theta=0.51$ which is optimized to maximize the storage capacity.
The storage capacity $\alpha_C$ is $0.44$.
Figure \ref{fig2}(b) shows the dependency of the overlap $m$ on the loading rate $\alpha$ with the synaptic depression, i.e., $\gamma=1.0$, $\tau=2.0$, $U_{SE}=0.5$, $x_j(0)=0.5$, and $\hat{\theta}=\frac{\theta}{1+\gamma}=0.255$.
The factor $x_j(t)$ obeys equation (\ref{eq.depression}) in computer simulations.
The storage capacity $\alpha_C$ is $0.44$.
This results show that the synaptic depression does not change the steady states.

\section{Basins of Attraction}
We investigate how the synaptic depression influences basins of attraction.
When a loading rate $\alpha$ is less than a storage capacity $\alpha_C$, a critical overlap $m_C$ exists \cite[]{Amari88}.
When an initial overlap $m^1(0)$ is larger than the critical overlap $m_C$, the retrieval of the memory pattern $\bm{\xi}^1$ succeeds.
In other words, the system converges to the pattern $\bm{\xi}^1$.
Therefore, the region of $m^1(0)>m_C$ is known as the basins of attraction.
The basins of attraction express an error-correcting ability of the model.
If the basins of attraction are enlarged, it means that the error-correcting ability of the associative memory model is improved.
In this section, we consider excitatory neurons, i.e., $g=0$, and this is same case as section 4.

Here, we investigate the basins of attraction without the synaptic depression, i.e., $\gamma=0$.
At first, we consider the case where the number of memory patterns is small, i.e., $p \sim O(1)$.
Let us consider the probability distribution of the internal potential $h_i(0)$ at time $t=0$.
The peak of the distribution of $\xi_i^1=1$ is located at $(1-f)m^1(0)$.
If the peak of the distribution of $\xi_i^1=1$ is smaller than the threshold $\theta$, the states of all neurons become $0$ at time $t=1$ and the retrieval fails (Figure \ref{fig3}(a)).
Therefore, the peak is larger than the threshold $\theta$, i.e., $(1-f)m^1(0) > \theta$ for the successful retrieval (Figure \ref{fig3}(b)).
When the loading rate $\alpha$ is small, the critical overlap $m_C$ satisfies the relationship of $m_C=\frac{\theta}{1-f}$.

Next, we consider the case where the number of memory patterns is order $N$, i.e., $p \sim O(N)$.
If the threshold is set at a small value at $t=0$, the initial overlap $m^1(0)$ can be a small value.
If the threshold is fixed at a small value, the threshold crosses the distribution of $\xi^1_i=0$ because of the cross-talk noise (Figure \ref{fig4}(a)).
At the next time $t=1$, the neuronal state whose internal potential $h_i(0)$ is larger than the threshold $\theta$ becomes $1$ even though the neuron codes $\xi^1_i=0$ (shadow part of Figure \ref{fig4}(a)).
The mean firing rate of the model increases, and the overlap decreases.
Then the distribution of $\xi^1_i=1$ is smaller than the threshold, and the retrieval fails.
If the threshold increases at $t=1$ to keep the mean firing rate at a constant level, the overlap increases at the next time $t=2$.
As the threshold increases to maintain the mean firing rate at a constant level, the overlap increases (Figure 4(b)).
This implies that if the threshold increases gradually for the increase of the signal in the progress of the retrieval, even though the initial overlap $m^1(0)$ is a small value, the retrieval succeeds.
In other words, the basins of attraction are enlarged.
This is an activity control mechanism \cite[]{Okada96,Matsumoto02}.

Here, we investigate the basins of attraction with the synaptic depression, i.e., $\gamma>0$.
We consider the case where the number of memory patterns is order $N$, i.e., $p \sim O(N)$.
We set $x_j(0)=1$, which implies that the synaptic depression does not work at time $t=0$.
Let the threshold $\hat{\theta}$ be fixed at a small value.
The distribution of the internal potential $h_i(0)$ at time $t=0$ is shown in Figure \ref{fig5}(a).
In the progress of the retrieval, the overlap increases.
This is a similar case as Figure \ref{fig4}(a).
By using the synaptic depression, the internal potential is written as
\begin{equation}
h_i(t) = \sum_{j=1}^N \tilde{J}_{ij} x_j(t) s_j(t).
\end{equation}
In the progress of the retrieval, $x_j(t)$ following equation (\ref{eq.depression}) decreases.
The signal might be fixed because the effect of the increase of the signal can be canceled out by the decrease of $x_j(t)$.
At a steady state, the distribution of $h_i$ is shown in Figure \ref{fig4}(b).
The relative relationship between the fixed threshold and the signal does not change in the progress of the retrieval.
In the activity control mechanism, the relative relationship between the threshold and the signal does not change in the progress of the retrieval.
Thus, the synaptic depression might have qualitatively same mechanism as the activity control mechanism.

In order to check the qualitative consideration, we calculate the basins of attraction by computer simulations.
Figure \ref{fig6}(a) shows the basins of attraction without the synaptic depression, i.e., $\gamma=0$.
The region in which $m^1(0)$ is larger than $m_C$ represents the basin of attraction for the retrieval of the target pattern $\bm{\xi}^1$.
The value $m_C$ is obtained by setting the initial state of the network at $\bm{\xi}^1$ with additional noise.
We employ the following method to add noise. 
$100y\%$ of the minority components ($s_i(0)=1$) are flipped, while the same number of majority components ($s_i(0)=0$) are also flipped.
The initial overlap $m^1(0)$ is given as $1-\frac{2y}{1-f}$.
Then the mean firing rate of the model is kept equal to the firing rate of the memory pattern, $f$.
When the threshold $\theta$ is fixed at an optimal value which maximize the storage capacity $\alpha_C$, the basins of attraction are small ($-$ in Figure \ref{fig6}(a)).
When the threshold $\theta$ is small, the basins of attraction are enlarged, but the storage capacity decreases ($\square$, $\times$, $*$ in Figure \ref{fig6}(a)).
Figure \ref{fig6}(b) shows the basins of attraction with the synaptic depression.
As discussed in section 4, the storage capacity $\alpha_C$ takes $0.44$ because the threshold is set at $\hat{\theta}=\theta/(1+\gamma)$.
Since the optimal threshold is $\theta=0.51$ without the synaptic depression, the threshold $\hat{\theta}$ with the synaptic depression is set at $\hat{\theta}=\theta/(1+\gamma)=0.425$ ($\gamma=0.2$,$\square$ in Figure \ref{fig6}(b)), $\hat{\theta}=0.340$ ($\gamma=0.5$, $\times$ in Figure \ref{fig6}(b)), $\hat{\theta}=0.255$ ($\gamma=1.0$, $*$ in Figure \ref{fig6}(b)).
As the value of $\gamma$ increases, the basins of attraction are more enlarged.

To check the qualitative consideration that the synaptic depression might have same mechanism as the activity control mechanism, we compare the temporal change of the overlap in the cases using the synaptic depression and the activity control mechanism.
Figure \ref{fig7}(a) and (b) shows the temporal change of the overlap $m^1(t)$ and the factor $x_j(t)$, respectively.
The parameters used in Figure \ref{fig7}(a) and (b) are optimized to enlarge the basins of attraction.
Figure \ref{fig7}(a) indicates that the system converges to the target pattern $\bm{\xi}^1$ within $4$ time steps.
Figure \ref{fig7}(b) indicates that the value $x_j(t)$ converges at $5$ time steps.
This result indicates that it is crucial that the time constant of the convergence to $\bm{\xi}^1$ is close to that of the convergence of $x_j(t)$ in order to enlarge the basins of attraction.
To compare the effect of the activity control mechanism with that of the synaptic depression, the temporal change of $m^1(t)$ with the activity control mechanism is shown in Figure \ref{fig7}(c).
The threshold is set to maintain the mean firing rate of the model at $f$.
Figure \ref{fig7}(c) indicates that the system converges to $\bm{\xi}^1$ within $5$ time steps.
The dynamics of the overlap is similar to when the synaptic depression is incorporated (figure \ref{fig7}(a)).
The basin of attraction in this case is larger than when the synaptic depression is incorporated.
The temporal change of the threshold is shown in figure \ref{fig7}(d).
The threshold converges at $5$ time steps.
These results support that the synaptic depression might have qualitively same mechanism as the activity control mechanism.

Bibitchkov et al. reported that the synaptic depression enlarged the basins of attraction a little at a small loading rate but it made the  basins of attraction shrink at a large loading rate \cite[]{Bibitchkov02}.
Setting a threshold at an appropriate value is critical in terms of increasing the storage capacity and enlarging the basins of attraction \cite[]{Okada96}.
However, Bibitchkov et al. did not set a threshold at an appropriate value.
Thus, they could not separate the effect of the synaptic depression and that of a threshold.
In contrast, we set a suitable threshold to avoid the effect of a threshold, and we can discuss only the effect of the synaptic depression.
Thus, we obtain the results that the synaptic depression enlarges the basins of attraction not only at a small loading rate but at a large loading rate.
The results obtained in this study are qualitatively different from the results obtained by \cite{Bibitchkov02}.

\section{Excitatory-Inhibitory Balanced Network}
In sections 4 and 5, we considered the model which consisted of excitatory neurons, i.e., $g=0$.
Here, we consider the excitatory-inhibitory balanced network, i.e., $g>0$.
It is known that inhibitory neurons regulate the overall activity of excitatory neurons.
If the overall activity of the excitatory neurons goes up, the inhibitory neurons send strong inhibition to the excitatory neurons to suppress the overall activity of the excitatory neurons.
If the overall activity of the excitatory neurons goes down, the inhibitory neurons become silent.
This excitatory-inhibitory balanced network must play the role of an activity control in cortical circuits.
In other words, the inhibitory neurons control the effective threshold of the excitatory neurons.
In section 5 we found that the synaptic depression might have qualitatively same mechanism as the activity control mechanism, even though the synaptic depression is a local phenomenon and the activity control is a global phenomenon.
In this section, we investigate whether the excitatory-inhibitory balance and the synaptic depression work cooperatively or not.

To do this, we consider the excitatory-inhibitory balanced network which does not incorporate the synaptic depression, i.e., $\gamma=0$.
The state of the $i$-th excitatory neuron is determined by
\begin{eqnarray}
s_i(t+1) &=& \Theta\Big( \sum^N_{j \ne i} \tilde{J}_{ij}s_j(t) - g\big(\bar{s}(t) -f \big) - \theta \Big)\\
         &=& \Theta\Big( \sum^N_{j \ne i} \tilde{J}_{ij}s_j(t) - \bar{\theta}(t) \Big),
\end{eqnarray}
where $\bar{s}(t)=\frac{1}{N}\sum^N_{j \ne i} s_j(t)$ and $\bar{\theta}(t)=g\big(\bar{s}(t) -f \big)+\theta$.
When the retrieval of memory patterns succeeds, the mean firing rate of the excitatory neurons, $\bar{s}(t)$, is close to $f$.
Therefore, the effect of the inhibition disappears:
\begin{equation}
g(\bar{s}(t) -f) \approx 0.
\end{equation}
This means that the storage capacity does not change.

We investigate whether the balanced network enlarged the basins of attraction.
At first, we consider the case where the number of memory patterns is order $1$, i.e., $p \sim O(1)$.
The cross-talk noise does not exist in this case.
When the initial overlap $m^1(0)$ is smaller than $\frac{\theta}{1-f}$, no neurons fire at the next time $t=1$.
Even if the effective threshold $\bar{\theta}(t)$ decreases at time $t=1$, the distribution of $\xi_i^1=1$ also decreases because the overlap takes a very small value.
Therefore, the retrieval fails.
In other words, the basins of attraction are enlarged only a little when $p$ is a small value.

Next, we consider the case where the number of memory patterns is order $N$, i.e., $p \sim O(N)$.
The distribution of an internal potential is shown in Figure \ref{fig8}.
When the initial overlap $m^1(0)$ is smaller than $\frac{\theta}{1-f}$, almost all neurons do not fire at the next time $t=1$.
However, at $p \sim O(N)$ each distribution becomes broader because of the cross-talk noise.
This enables some neurons to activate at time $t=1$.
Since the mean firing rate of the excitatory neurons, $\bar{s}(t)$, is smaller than $f$, the effective threshold $\bar{\theta}(t)$ decreases like Figure \ref{fig8}(b).
In the progress of the retrieval, the effective threshold changes over time, following the value of $\bar{s}(t)$.
If the model retrieves the target pattern $\bm{\xi}^1$, $\bar{s}(t)$ is close to $f$.
In other words, the inhibition disappears at the steady state (Figure \ref{fig8}(c)).

Here, the synaptic depression is incorporated into the excitatory-inhibitory balanced network, i.e., $\gamma>0$.
For simplicity, we assume that the synaptic depression is occurred on the synaptic sites among the excitatory neurons.
Then the state of the $i$-th excitatory neuron can be written as
\begin{equation}
s_i(t+1) = \Theta\Big( \sum^N_{j \ne i} \tilde{J}_{ij}x_j(t)s_j(t) - g\big(\bar{s}(t) -f \big) - \hat{\theta} \Big).
\end{equation}
We consider the case where the number of memory patterns is order $1$, i.e., $p \sim O(1)$.
The retrieval fails at an initial overlap $m^1(0)$ without the synaptic depression when $(1-f)m^1(0)$ is lower than the threshold $\theta$ (Figure \ref{fig9}(a)).
On the other hand, the retrieval succeeds at the same initial overlap $m^1(0)$ with the synaptic depression (Figure \ref{fig9}(b)) because $(1-f)m^1(0)$ is higher than the threshold $\hat{\theta}=\frac{\theta}{1+\gamma}$.
Therefore, the basins of attraction are enlarged at a small loading rate.
At a large loading rate ($p \sim O(N)$), the effect of the synaptic depression is smaller than that of the inhibition because of the cross-talk noise.
Thus, at a large loading rate, the synaptic depression does not change the basins of attraction in the excitatory-inhibitory balanced network.

In order to check the qualitative consideration, we calculate the basins of attraction by computer simulations.
Figure \ref{fig10}(a) shows the basins of attraction without the synaptic depression.
As the value of $g$ increases, the basins of attraction are enlarged.
At a small loading rate the basins are enlarged a little.
Thus, the excitatory-inhibitory balanced network enlarges the basins of attraction. 
At a small loading rate it enlarges the basins a little.
Figure \ref{fig10}(b) shows the basins of attraction with the synaptic depression.
To compare the effect of the synaptic depression, the case of the balanced network without the synaptic depression ($\square$) is shown.
In the balanced network with the synaptic depression, the basins of attraction are the largest ($*$).
Even at a small loading rate the basins are enlarged.
Thus, the excitatory-inhibitory balance and the synaptic depression work cooperatively.

\section{Discussion}
In this paper, we investigated how the synaptic depression influenced the performance of the associative memory model in terms of the storage capacity and the basins of attraction.
Using the mean-field theory and the computer simulations, we found that the basins of attraction were enlarged whereas the storage capacity did not decrease.
In other words, the synaptic depression had the mechanisms that the neuron threshold effectively increased in the progress of the retrieval.
Furthermore, the excitatory-inhibitory balance and the synaptic depression worked cooperatively in the excitatory-inhibitory balanced network.
This result suggests that the short-term synaptic depression might improve an error-correcting ability in cortical circuits.

In our model the synaptic depression was assumed to occur only at the synaptic sites among excitatory neurons.
In the cortical circuits the synaptic depression occurs not only at excitatory synaptic sites but at inhibitory synaptic sites.
Recently, Galarreta and Hestrin found that long-term firing induced a much stronger depression of excitatory synapses than that of inhibitory synapses although the initial rates of depression were similar at both excitatory and inhibitory synapses \cite[]{Galarreta98}.
The result implies that the synaptic depression induces strong depression at excitatory synapses, it induces weak depression at inhibitory synapses.
If the weak depression at inhibitory synapses is regarded as no depression in our model, our assumption that the synaptic depression occurs among the excitatory neurons might be valid.

\newpage
\bibliographystyle{apalike}

\begin{thebibliography}{}

\bibitem[Abbott et~al., 1997]{Abbott97}
Abbott, L., Varela, J., Sen, K., and Nelson, S. (1997).
\newblock Synaptic depression and cortical gain control.
\newblock {\em Science}, 275:220--224.

\bibitem[Amari and Maginu, 1988]{Amari88}
Amari, S. and Maginu, K. (1988).
\newblock Statistical neurodynamics of various versions of correlation
  associative memory.
\newblock {\em Neural Network}, 1:63--73.

\bibitem[Amit, 1989]{Amit89}
Amit, D. (1989).
\newblock {\em Modeling brain function: the world of attractor neural
  networks}.
\newblock Cambridge university press.

\bibitem[Amit et~al., 1994]{Amit94}
Amit, D., Brunel, N., and Tsodyks, M. (1994).
\newblock Correlations of cortical hebbian reverberations: theory versus
  experiment.
\newblock {\em Journal of Neuroscience}, 14:6435--6445.

\bibitem[Bibitchkov et~al., 2002]{Bibitchkov02}
Bibitchkov, D., Herrmann, J., and Geisel, T. (2002).
\newblock Pattern storage and processing in attractor networks with short-time
  synaptic dynamics.
\newblock {\em Network: Computation in Neural Systems}, 13:115--129.

\bibitem[Bressloff, 1999]{Bressloff99}
Bressloff, P. (1999).
\newblock Dynamic synapses, a new concept of neural representation and
  computation.
\newblock {\em Physical Review E}, 60:2160--2170.

\bibitem[Galarreta and Hestrin, 1998]{Galarreta98}
Galarreta, M. and Hestrin, S. (1998).
\newblock Frequency-dependent synaptic depression and the balance of excitation
  and inhibition in the neocortex.
\newblock {\em Nature Neuroscience}, 1:587--594.

\bibitem[Hebb, 1949]{Hebb49}
Hebb, D. (1949).
\newblock {\em The Organization of Behavior: A Neuropsychological Theory}.
\newblock Wiley, New York.

\bibitem[Kistler and Hemmen, 1999]{Kistler99}
Kistler, W. and Hemmen, J. (1999).
\newblock Short-term synaptic plasticity and network behavior.
\newblock {\em Neural Computation}, 11:1579--1594.

\bibitem[Matsumoto and Okada, 2002]{Matsumoto02}
Matsumoto, N. and Okada, M. (2002).
\newblock Self-regulation mechanism of temporally asymmetric hebbian
  plasticity.
\newblock {\em Neural Computation}, 14:2883--2902.

\bibitem[Matsumoto et~al., 2005]{Matsumoto05}
Matsumoto, N., Okada, M., Sugase-Miyamoto, Y., and Yamane, S. (2005).
\newblock Neuronal mechanisms encoding global-to-fine information in
  inferior-temporal cortex.
\newblock {\em Journal of Computational Neuroscience}, 18:85--103.

\bibitem[Okada, 1996]{Okada96}
Okada, M. (1996).
\newblock Notions of associative memory and sparse coding.
\newblock {\em Neural Networks}, 9:1429--1458.

\bibitem[Pantic et~al., 2002]{Pantic02}
Pantic, L., Torres, J., Kappen, H., and Gielen, S. (2002).
\newblock Associative memory with dynamic synapses.
\newblock {\em Neural Computation}, 14:2903--2923.

\bibitem[Shiino and Fukai, 1992]{Shiino92}
Shiino, M. and Fukai, T. (1992).
\newblock Self-consistent signal-to-noise analysis and its application to
  analogue neural networks with asymmetric connections.
\newblock {\em Journal of Physics A: Mathematical and General}, 25:L375--L381.

\bibitem[Thomson and Deuchars, 1994]{Thomson94}
Thomson, A. and Deuchars, J. (1994).
\newblock Temporal and spatial properties of local circuits in neocortex.
\newblock {\em Trends in Neuroscience}, 17:119--126.

\bibitem[Torres et~al., 2002]{Torres02}
Torres, J., Pantic, L., and Kappen, H. (2002).
\newblock Storage capacity of attractor neural networks with depressing
  synapses.
\newblock {\em Physical Review E}, 66:061910--1--5.

\bibitem[Tsodyks and Markram, 1997]{Tsodyks97}
Tsodyks, M. and Markram, H. (1997).
\newblock The neural code between neocortical pyramidal neurons depends on
  neurotransmitter release probability.
\newblock {\em Proceedings of the National Academy of Sciences}, 94:719--723.

\bibitem[Tsodyks et~al., 2000]{Tsodyks00}
Tsodyks, M., Uziel, A., and Markram, H. (2000).
\newblock Synchrony generation in recurrent networks with frequency-dependent
  synapses.
\newblock {\em Journal of Neuroscience}, 20:RC50--1--5.

\bibitem[Vreeswijk and Sompolinsky, 1996]{Vreeswijk96}
Vreeswijk, C. and Sompolinsky, H. (1996).
\newblock Chaos in neuronal networks with balanced excitatory and inhibitory
  activity.
\newblock {\em Science}, 274:1724--1726.

\end{thebibliography}

\clearpage
\appendix
\section{Derivation of Mean-Field Equations}
In this appendix, we derive mean-field equations at a steady state, i.e., $t \rightarrow \infty$.
For simplicity, each value at the steady state is written by $x_j(\infty)=x_j$, $h_i(\infty)=h_i$, $m^{\mu}(\infty)=m^{\mu}$, and $s_i(\infty)=s_i$.
The factor $x_j(t)$ reaches its steady-state value by $t \rightarrow \infty$ in equation (\ref{eq.depression}) \cite[]{Bibitchkov02}:
\begin{equation}
x_j=\frac{1}{1+\gamma s_j},
\end{equation}
where $\gamma=\tau U_{SE}$.
Since $s_j$ takes binary values, i.e., $s_j = \{0,1\}$, $x_j s_j$ is written as
\begin{equation}
x_j s_j=\frac{s_j}{1+\gamma s_j}=\frac{1}{1+\gamma}s_j.
\end{equation}
By using this relationship, the state of $i$-th excitatory neuron is written as
\begin{eqnarray}
s_i &=& \Theta\Big( \sum_{j \ne i}^N \tilde{J}_{ij}x_js_j -g(\bar{s} - f) -\hat{\theta}\Big)\\
&=& \Theta\Big( \frac{1}{1+\gamma}\sum_{j \ne i}^N \tilde{J}_{ij}s_j -g(\bar{s} - f) -\hat{\theta}\Big)\\
&=& \Theta\Big(\frac{1}{1+\gamma}\big(\sum_{j \ne i}^N \tilde{J}_{ij}s_j -g(\bar{s} - f)- (1+\gamma)\hat{\theta}\big)\Big)\\
&=& \Theta\Big( \sum_{j \ne i}^N \tilde{J}_{ij}s_j -(1+\gamma)g(\bar{s} - f)-(1+\gamma)\hat{\theta}\Big).
\end{eqnarray}
where $\tilde{J}_{ij}=\frac{1}{Nf(1-f)}\sum^p_{\mu =1}(\xi^{\mu}_i-f)(\xi^{\mu}_j-f)$ and $\hat{\theta}$ indicates a threshold when the synaptic depression is incorporated into the model. 
The neuronal state can be transformed by
\begin{eqnarray}
s_i &=& \Theta \Big( \frac{1}{Nf(1-f)}\sum^p_{\mu =1}\sum^N_{j \ne i}(\xi^{\mu}_i-f)(\xi^{\mu}_j-f)s_j -(1+\gamma)g(\bar{s} - f)- (1+\gamma)\hat{\theta} \Big) \nonumber \\ \\
&=& \Theta \Big(\frac{1}{Nf(1-f)}\sum^p_{\mu =1}\sum^N_{j=1}(\xi^{\mu}_i-f)(\xi^{\mu}_j-f)s_j\nonumber \\
& & - \frac{1}{Nf(1-f)} \sum^p_{\mu=1} (\xi^{\mu}_i-f)(\xi^{\mu}_i-f)s_i -(1+\gamma)g(\bar{s} - f)- (1+\gamma)\hat{\theta} \Big)\\
&=& \Theta \Big(\sum_{\mu=1}^p (\xi^{\mu}_i -f)m^{\mu} - \alpha s_i -(1+\gamma)g(\bar{s} - f)- (1+\gamma)\hat{\theta} \Big) \label{eq.self}.
\end{eqnarray}
We assume that equation (\ref{eq.self}) can be solved by using an effective response function $F(\cdot)$ as
\begin{equation}
s_i = F \Big(\sum_{\mu=1}^p (\xi^{\mu}_i -f)m^{\mu} -(1+\gamma)g(\bar{s} - f)- (1+\gamma)\hat{\theta}\Big).
\end{equation}
Let $\bm{\xi}^{1}$ be a target pattern to be retrieved.
We assume that $m^{1}$ is order of $1$ with respect to $N$ ($m^1 \sim O(1)$) and $m^{\mu}$ ($\mu > 1$) is order of $\frac{1}{\sqrt{N}}$ with respect to $N$ ($m^{\mu} \sim O(\frac{1}{\sqrt{N}}))$.
We apply Taylor expansion to obtain
\begin{eqnarray}
s_i &=& F \Big((\xi^{\mu}_i-f)m^{\mu} + \sum_{\nu \neq \mu}^p (\xi^{\nu}_i -f)m^{\nu} -(1+\gamma)g(\bar{s} - f)- (1+\gamma)\hat{\theta}\Big)\\
&=& F \Big(\sum_{\nu \neq \mu}^p (\xi^{\nu}_i -f)m^{\nu} -(1+\gamma)g(\bar{s} - f)- (1+\gamma)\hat{\theta}\Big) \nonumber\\
& &+ (\xi^{\mu}_i-f)m^{\mu} F'\Big( \sum_{\nu \neq \mu}^p (\xi^{\nu}_i -f)m^{\nu} -(1+\gamma)g(\bar{s} - f)- (1+\gamma)\hat{\theta}\Big) \\
&=& s_i^{(\mu)} + (\xi^{\mu}_i-f)m^{\mu} s_i'^{(\mu)},
\end{eqnarray}
where $s_i^{(\mu)}=F \Big(\sum_{\nu \neq \mu}^p (\xi^{\nu}_i -f)m^{\nu} -(1+\gamma)g(\bar{s} - f)- (1+\gamma)\hat{\theta}\Big)$ and $s_i'^{(\mu)}=F'\Big(\sum_{\nu \neq \mu}^p (\xi^{\nu}_i -f)m^{\nu} -(1+\gamma)g(\bar{s} - f)- (1+\gamma)\hat{\theta}\Big)$.
By using this relationship, an overlap $m^{\mu}$ is obtained as 
\begin{eqnarray}
m^{\mu} &=& \frac{1}{Nf(1-f)}\sum^N_{i=1}(\xi_i^{\mu}-f) s_i\\
&=& \frac{1}{Nf(1-f)}\sum^N_{i=1}(\xi_i^{\mu}-f)s_i^{(\mu)} + \frac{1}{Nf(1-f)}\sum^N_{i=1}(\xi_i^{\mu}-f)^2 m^{\mu} s_i'^{(\mu)} \\
&=& \frac{1}{Nf(1-f)}\sum^N_{i=1}(\xi_i^{\mu}-f)s_i^{(\mu)} + Um^{\mu},
\end{eqnarray}
where $U = \frac{1}{N} \sum^N_{i=1} s_i'^{(\mu)}$.
The overlap $m^{\mu}$ is obtained as
\begin{equation}
m^{\mu}=\frac{1}{Nf(1-f)}\frac{1}{1-U}\sum^N_{i=1}(\xi^{\mu}_i-f)s^{(\mu)}_i.
\end{equation}
The first term of equation (39) is transformed by
\begin{eqnarray}
\sum^N_{j \ne i}\tilde{J}_{ij}s_j &=& \frac{1}{Nf(1-f)}\sum^p_{\mu =1}\sum^N_{j \ne i}(\xi^{\mu}_i-f)(\xi^{\mu}_j-f)s_j\\
&=&(\xi_i^1-f)m^1 + \sum^p_{\mu=2}(\xi^{\mu}_i-f)m^{\mu} - \alpha s_i\\
&=&(\xi_i^1-f)m^1 + \frac{1}{Nf(1-f)}\frac{1}{1-U}\sum^p_{\mu=2}\sum^N_{j=1}(\xi^{\mu}_i-f)(\xi^{\mu}_j-f)s^{(\mu)}_j - \alpha s_i \\
&=& (\xi_i^1-f)m^1 +\frac{1}{Nf(1-f)}\frac{1}{1-U}\sum^p_{\mu=2}\sum^N_{j \ne i}(\xi^{\mu}_i-f)(\xi^{\mu}_j-f)s_j^{(\mu)}+\frac{\alpha U}{1-U}s_i \nonumber \\ \\
&=& (\xi_i^1-f)m^1 + \tilde{z}_i + \Gamma s_i
\end{eqnarray}
where $\tilde{z}_i=\frac{1}{Nf(1-f)}\frac{1}{1-U}\sum^p_{\mu=2}\sum^N_{j \ne i}(\xi^{\mu}_i-f)(\xi^{\mu}_j-f)s_j^{(\mu)}$ and $\Gamma=\frac{\alpha U}{1-U}$.
In this transformation we used $s_i=s_i^{(\mu)}$ because of $s_i-s_i^{(\mu)} \sim O(\frac{1}{\sqrt{N}})$.
The first term in equation (55) is a signal term for the retrieval of the target pattern $\bm{\xi}^1$ while the second term is a cross-talk noise.
Since $s_j^{(\mu)}$ is independent of $\xi_j^{\mu}$, $\tilde{z}_i$ obeys a Gaussian distribution whose average is $0$ and variance is $\sigma^2$, i.e. $\tilde{z}_i \sim N(0,\sigma^2)$.
Since the square of $\tilde{z}_i$ is calculated by
\begin{equation}
(\tilde{z}_i)^2 = \Big( \frac{1}{Nf(1-f)}\frac{1}{1-U}\sum^p_{\mu=2}\sum^N_{j \ne i}(\xi^{\mu}_i-f)(\xi^{\mu}_j-f)s_j^{(\mu)} \Big)^2 = \frac{\alpha q}{(1-U)^2},
\end{equation}
the variance $\sigma^2$ is calculated by
\begin{equation}
\sigma^2 = \mbox{E}[(\tilde{z}_i)^2] - \Big( \mbox{E}[\tilde{z}_i] \Big)^2 = \frac{\alpha q}{(1-U)^2},
\end{equation}
where $q = \frac{1}{N}\sum^N_{j=1}(s^{(\mu)}_j)^2$.
The third term in equation (55) is an average term of the cross-talk noise and denotes an effective self-coupling term.
Since $s_i$ is symmetric about $i$, $s_i$ can be replaced by $Y$:
\begin{equation}
Y = \Theta \Big((\xi_i^1-f)m^1 + \tilde{z}_i + \Gamma Y -(1+\gamma)g(\bar{s} - f)- (1+\gamma)\hat{\theta} \Big).
\end{equation}
By using a random variable $z_i \sim N(0,1)$, $Y$ is written as
\begin{equation}
Y = \Theta \Big((\xi_i^1-f)m^1 + \sigma z_i + \Gamma Y -(1+\gamma)g(\bar{s} - f)- (1+\gamma)\hat{\theta} \Big).
\end{equation}
We assume a self-averaging property to hold so that a site average can be replaced by an average over the memory patterns and random variable $z_i \sim N(0,1)$ in which the random variation of $z_i$ from site to site can be described by a single noise $z$ obeying an identical Gaussian distribution.
After rewriting $\xi_i^1 \rightarrow \xi$, $m^1 \rightarrow m$, and $s_i \rightarrow Y$, and using the relationship of $s_i=s_i^{(\mu)}$ because of $s_i-s_i^{(\mu)} \sim O(\frac{1}{\sqrt{N}})$, the order-parameter equations are obtained as
\begin{eqnarray}
m &=&\frac{1}{f(1-f)}\int Dz <<(\xi-f) Y>>,\\
q &=&\int Dz <<Y^2>>,\\
U &=&\frac{1}{\sigma}\int Dzz <<Y>>,\\
\bar{s} &=& \int Dz <<Y>>, \\
Y &=& \Theta \Big((\xi-f)m + \sigma z + \Gamma Y -(1+\gamma)g(\bar{s} - f) - (1+\gamma)\hat{\theta} \Big), \\
\sigma^2 &=& \frac{\alpha q}{(1-U)^2}.
\end{eqnarray}
where $<<\cdots>>$ denotes the average over the memory pattern $\xi$ and random variable $z$ with the probability density $Dz = \frac{1}{\sqrt{2 \pi}}\exp (-\frac{z^2}{2})$.
In general, the equation of $Y$ (equation (64)) admits more then one solution.
In that case, Maxwell rule should be applied to choose the appropriate solution.
The solution $Y$ is given by
\begin{equation}
Y = \Theta \Big((\xi-f)m + \sigma z + \frac{\Gamma}{2} -(1+\gamma)g(\bar{s} - f)- (1+\gamma)\hat{\theta} \Big).
\end{equation}
Finally, we get the mean-field equations.
\begin{eqnarray}
m &=& \frac{1}{f(1-f)}\int Dz <<(\xi -f)Y>> \\ 
&=& \frac{1}{f(1-f)}\frac{1}{\sqrt{2 \pi}} \int \exp (-\frac{z^2}{2}) <<(\xi-f) \Theta\Big((\xi-f)m + \sigma z + \frac{\Gamma}{2} \nonumber \\& &-(1+\gamma)g(\bar{s} - f)- (1+\gamma)\hat{\theta} \Big)>> \\
&=& -\frac{1}{2} \mathrm{erf}(\phi_1) + \frac{1}{2} \mathrm{erf}(\phi_2),
\end{eqnarray}
where $\mathrm{erf}(y) = \frac{2}{\sqrt{\pi}}\int_0^y\mathrm{exp}(-u^2)du$, $\phi_1=\frac{-(1-f)m+(1+\gamma)g(\bar{s} - f)+(1+\gamma)\hat{\theta}-\frac{\Gamma}{2}}{\sqrt{2 \alpha \sigma^2}}$, and $\phi_2=\frac{fm+(1+\gamma)g(\bar{s} - f)+(1+\gamma)\hat{\theta}-\frac{\Gamma}{2}}{\sqrt{2 \alpha \sigma^2}}$.
\begin{eqnarray}
q &=& \int Dz <<Y^2>> \\
&=& \frac{1}{\sqrt{2 \pi}} \int \exp (-\frac{z^2}{2}) <<\Big\{ \Theta\Big((\xi-f)m + \sigma z + \frac{\Gamma}{2} -(1+\gamma)g(\bar{s} - f)- (1+\gamma)\hat{\theta} \Big) \Big\}^2>> \nonumber \\\\
&=& \frac{1}{2}-\frac{f}{2}\mathrm{erf}(\phi_1) - \frac{1-f}{2}\mathrm{erf}(\phi_2).
\end{eqnarray}
\begin{eqnarray}
U &=& \frac{1}{\sigma}\int Dz z<<Y>> \\
&=& \frac{1}{\sqrt{2 \pi} \sigma} \int z \exp (-\frac{z^2}{2}) <<\Theta\Big((\xi-f)m + \sigma z + \frac{\Gamma}{2} -(1+\gamma)g(\bar{s} - f)- (1+\gamma)\hat{\theta} \Big)>> \nonumber \\ \\
&=& \frac{f}{\sqrt{2\pi} \sigma}\exp(-\phi_1^2) + \frac{1-f}{\sqrt{2\pi} \sigma}\exp(-\phi_2^2),
\end{eqnarray}
\begin{eqnarray}
\bar{s} &=& \int Dz <<Y>> \\
&=& \frac{1}{\sqrt{2 \pi}} \int \exp (-\frac{z^2}{2}) << \Theta\Big((\xi-f)m + \sigma z + \frac{\Gamma}{2} -(1+\gamma)g(\bar{s} - f)- (1+\gamma)\hat{\theta} \Big)>> \nonumber \\\\
&=& \frac{1}{2}-\frac{f}{2}\mathrm{erf}(\phi_1) - \frac{1-f}{2}\mathrm{erf}(\phi_2).
\end{eqnarray}

\clearpage
\clearpage
\section*{Figure captions}
\begin{figure}[htb]
\caption{\label{fig1}
The probability distribution of $h_i$ at the steady state.
(a):$p \sim O(1)$ and the synapses are not depressed.
(b):$p \sim O(N)$ and the synapses are not depressed.
(c):$p \sim O(N)$ and the synapses are depressed.
}
\end{figure}

\begin{figure}[htb]
\caption{\label{fig2}
The dependency of $m$ on a loading rate $\alpha$ at $f=0.1$. 
The dashed lines are obtained by solving the mean-field equations numerically while the error bars indicate medians and quartile deviations of $m(100)$ obtained by computer simulations in $11$ trials at $N=5000$.
(a):the case without the synaptic depression, i.e., $\gamma=0$, and $\theta=0.51$. $\alpha_C=0.44$.
(b):the case with the synaptic depression, i.e., $\gamma=1.0$, $\tau=2.0$, $U_{SE}=0.5$, $x_j(0)=0.5$, and $\hat{\theta}=0.255$. $\alpha_C=0.44$.
}
\end{figure}

\begin{figure}[htb]
\caption{\label{fig3}
The probability distribution of $h_i(0)$.
$p \sim O(1)$ and the synapses are not depressed.
(a):the peak of the distribution of $\xi_i^1=1$ is smaller than the threshold $\theta$.
(b):the peak of the distribution of $\xi_i^1=1$ is larger than the threshold $\theta$.
}
\end{figure}

\begin{figure}[htb]
\caption{\label{fig4}
The probability distribution of the internal potential.
$p \sim O(N)$ and the synapses are not depressed.
(a):the threshold $\theta$ is fixed at a small value.
(b):the threshold $\theta(t)$ increases for the increase of the signal.
}
\end{figure}

\begin{figure}[htb]
\caption{\label{fig5}
The probability distribution of the internal potential.
$p \sim O(N)$ and the synapses are depressed.
(a):at time $t=0$.
(b):at the steady state.
}
\end{figure}

\begin{figure}[htb]
\caption{\label{fig6}
The dependency of $m_C$ on $\alpha$ at $f=0.1$. 
The dashed lines are obtained by solving the mean-field equations (same solid line in Figure 2) while the error bars indicate medians and quartile deviations obtained by computer simulations in $11$ trials at $N=5000$.
(a):the synapses are not depressed and $-$:$\theta=0.51$, $\square$:$\theta=0.425$, $\times$:$\theta=0.365$ and $*$:$\theta=0.255$.
(b):the synapses are depressed and $x_j(0)=1.0$. 
$-$:$\hat{\theta}=0.51$, $\gamma=0.0$.
$\square$:$\hat{\theta}=0.425$, $\gamma=0.2$, $\tau=1.2$, and $U_{SE}=0.167$.
$\times$:$\hat{\theta}=0.340$, $\gamma=0.5$, $\tau=1.5$, and $U_{SE}=0.333$.
$*$:$\hat{\theta}=0.255$, $\gamma=1.0$, $\tau=2.0$, and $U_{SE}=0.5$.
}
\end{figure}

\begin{figure}[htb]
\caption{\label{fig7}
(a):the temporal change of $m^1(t)$ with the synaptic depression.
Each solid line shows the temporal change of $m^1(t)$ when the initial overlap $m^1(0)$ is set at $0.1$, $0.2$, $0.3$, $0.4$, $0.5$, $0.6$, $0.7$, $0.8$, $0.9$, and $1.0$, respectively.
These lines were obtained by computer simulations at $\tau=2.5$, $U_{SE}=0.2$, $x_j(0)=1.0$, $p=500$, and $N=5000$.
(b):the temporal change of $x_j(t)$ when the initial value $x_j(0)$ is set at $0$, and the line is obtained by equation (8) at $\tau=2.5$, $U_{SE}=0.2$, and $x_j(0)=1.0$.
(c):the temporal change of $m^1(t)$ with the activity control mechanism.
Each solid line shows the temporal change of $m^1(t)$ when the initial overlap $m^1(0)$ is set at $0.1$, $0.2$, $0.3$, $0.4$, $0.5$, $0.6$, $0.7$, $0.8$, $0.9$, and $1.0$, respectively.
These lines were obtained by computer simulations at $p=500$ and $N=5000$.
(d):the temporal change of $\theta(t)$ when the initial overlap $m^1(0)$ is set at $0.2$.
}
\end{figure}

\begin{figure}[htb]
\caption{\label{fig8}
The distributions of the internal potential when the synapses are not depressed.
(a):at $t=0$, (b):at $t=1$, (c):at steady state.
}
\end{figure}

\begin{figure}[htb]
\caption{\label{fig9}
The probability distribution of $h_i(0)$ at time $t=0$ and $p \sim O(1)$.
(a):the synapses are not depressed, and the peak of the distribution $\xi_i^1=1$ is smaller than the threshold $\theta$.
(b):the synapses are depressed, and the peak of the distribution $\xi_i^1=1$ is larger than the threshold $\hat{\theta}$ at $x_j(0)=1.0$.
}
\end{figure}

\begin{figure}[htb]
\caption{\label{fig10}
The basins of attraction in the excitatory-inhibitory balanced network.
The dashed lines are obtained by the mean-field equations while the error bars indicate medians and quartile deviations obtained by computer simulations in $11$ trials at $N=5000$ and $f=0.1$.
(a):without the synaptic depression at $\theta=0.51$. $-$:$g=0$, $\times$:$g=1$, $*$:$g=2$, $\square$:$g=4.5$.
(b)$*$:with the synaptic depression ($\hat{\theta}=0.425$, $\gamma=0.2$, $g=4.5$)
$\square$:without the synaptic depression ($\theta=0.51$, $g=4.5$).
}
\end{figure}

\clearpage
\begin{figure}[p]
\begin{center}
\includegraphics{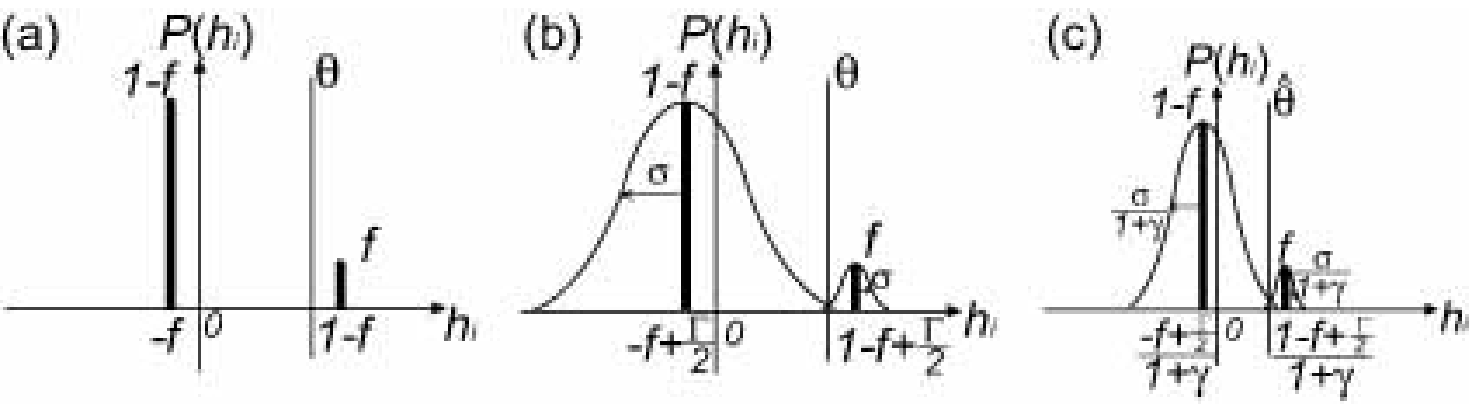}\\
\vspace{1.5cm}
{\Huge Figure 1}\\
\vspace{1cm}
\Large Matsumoto
\end{center}
\end{figure}

\clearpage
\begin{figure}[p]
\begin{center}
\includegraphics{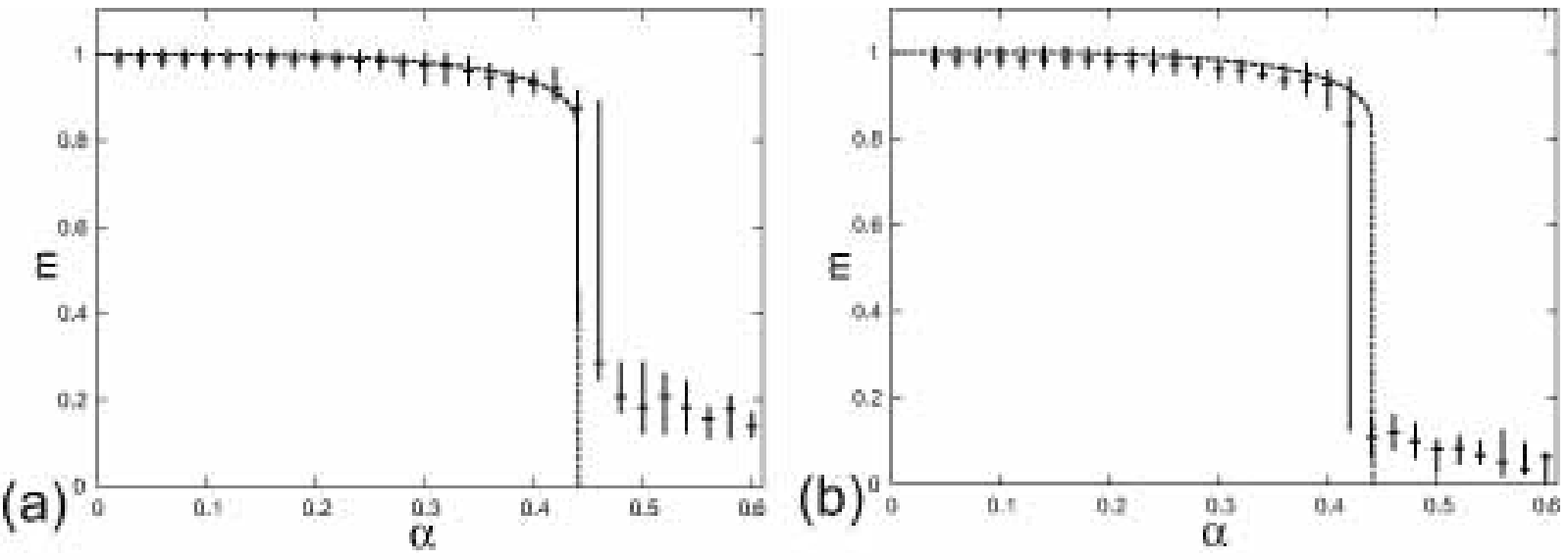}\\
\vspace{1.5cm}
{\Huge Figure 2}\\
\vspace{1cm}
\Large Matsumoto
\end{center}
\end{figure}

\clearpage
\begin{figure}[p]
\begin{center}
\includegraphics{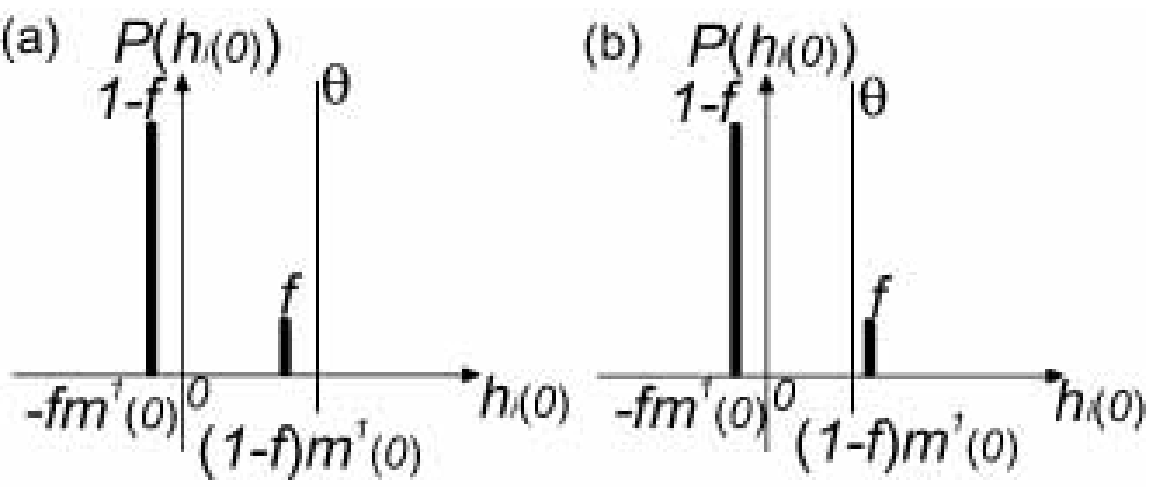}\\
\vspace{1.5cm}
{\Huge Figure 3}\\
\vspace{1cm}
\Large Matsumoto
\end{center}
\end{figure}

\clearpage
\begin{figure}[p]
\begin{center}
\includegraphics{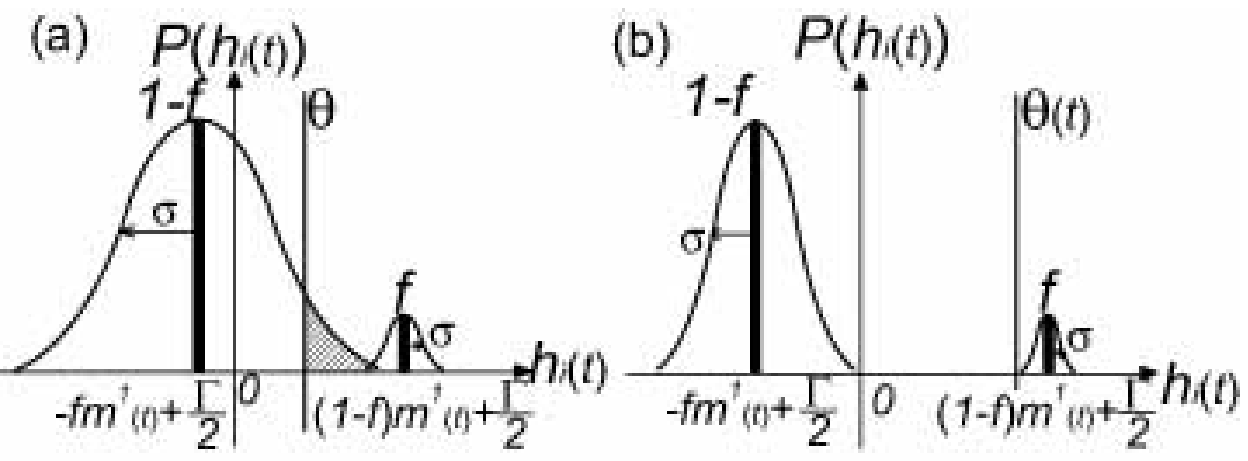}\\
\vspace{1.5cm}
{\Huge Figure 4}\\
\vspace{1cm}
\Large Matsumoto
\end{center}
\end{figure}

\clearpage
\begin{figure}[p]
\begin{center}
\includegraphics{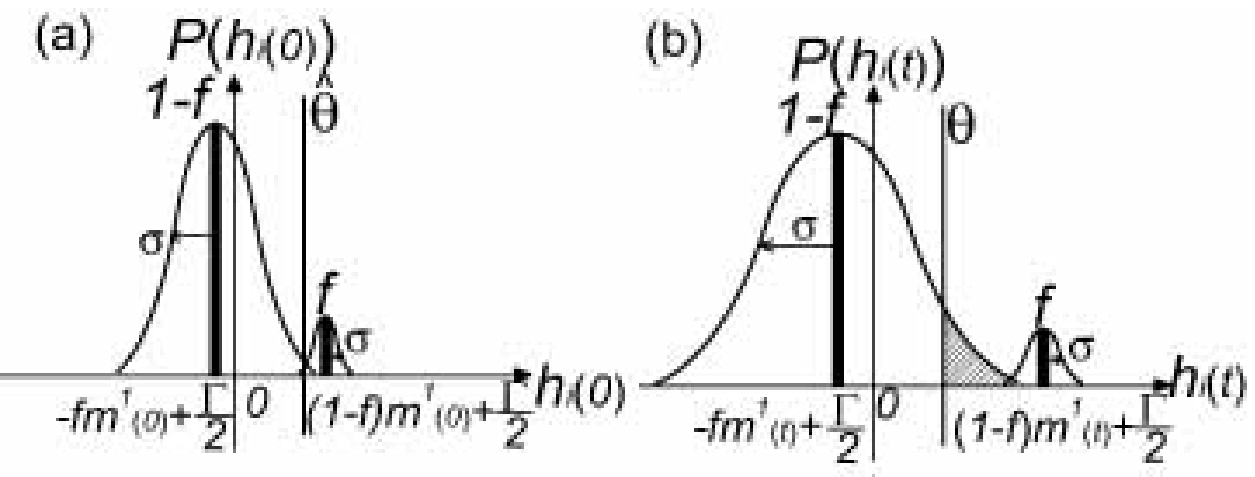}\\
\vspace{1.5cm}
{\Huge Figure 5}\\
\vspace{1cm}
\Large Matsumoto
\end{center}
\end{figure}

\clearpage
\begin{figure}[p]
\begin{center}
\includegraphics{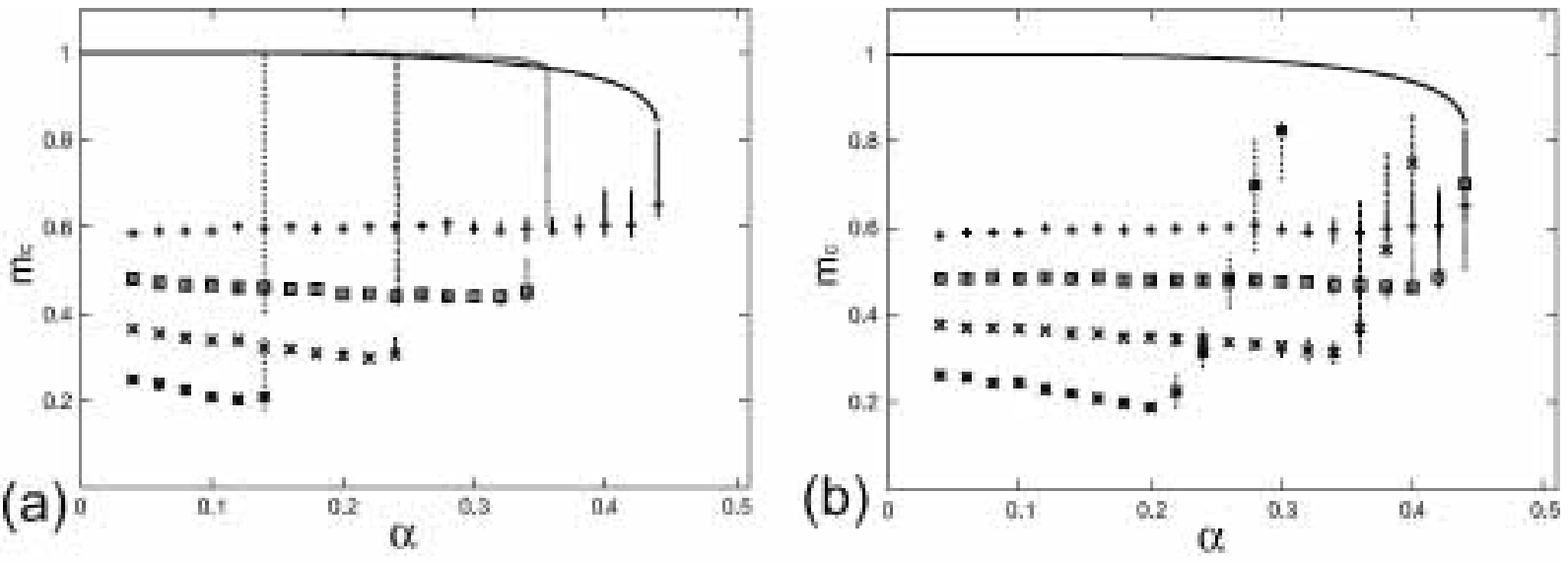}\\
\vspace{1.5cm}
{\Huge Figure 6}\\
\vspace{1cm}
\Large Matsumoto
\end{center}
\end{figure}

\clearpage
\begin{figure}[p]
\begin{center}
\includegraphics{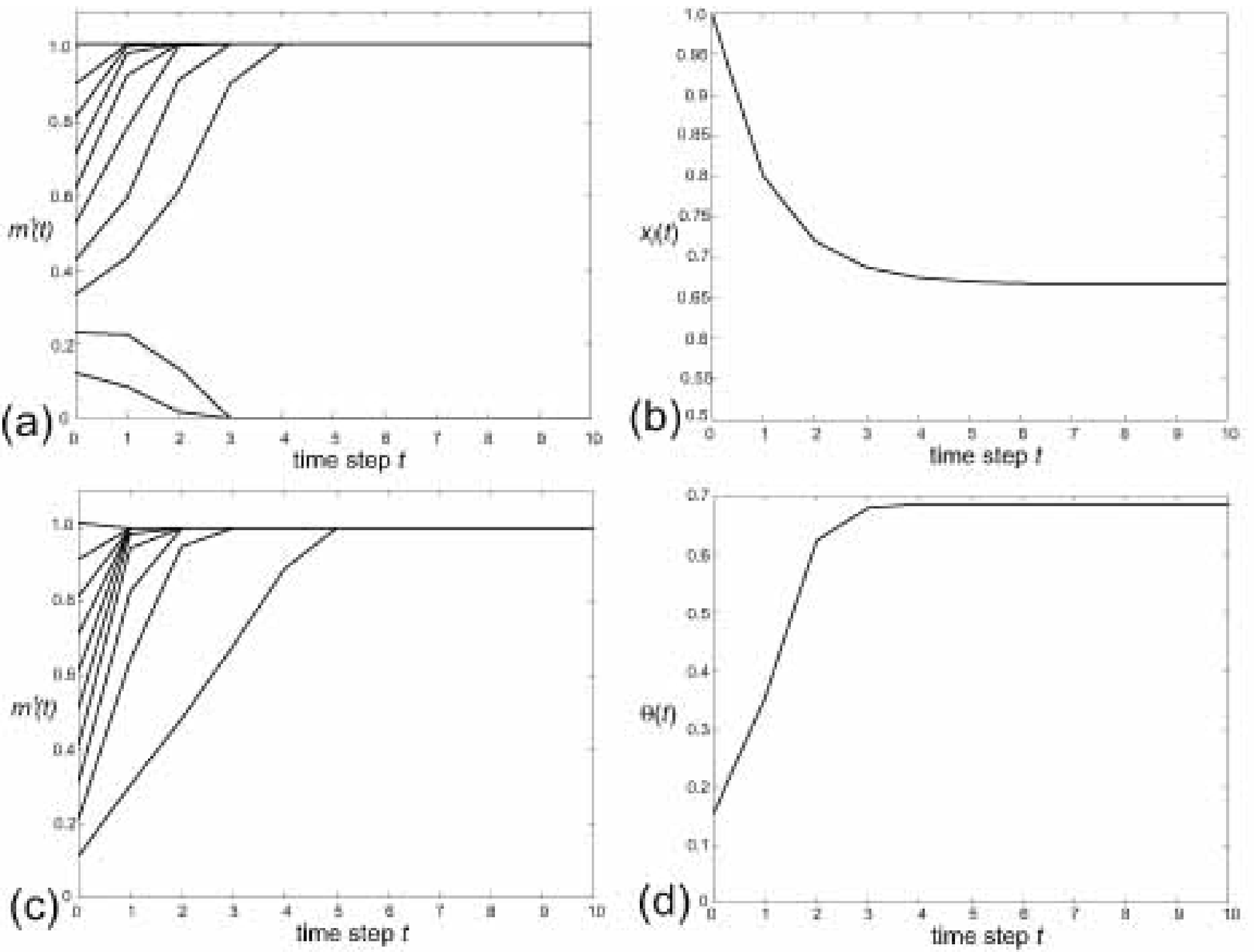}\\
\vspace{1.5cm}
{\Huge Figure 7}\\
\vspace{1cm}
\Large Matsumoto
\end{center}
\end{figure}

\clearpage
\begin{figure}[p]
\begin{center}
\includegraphics{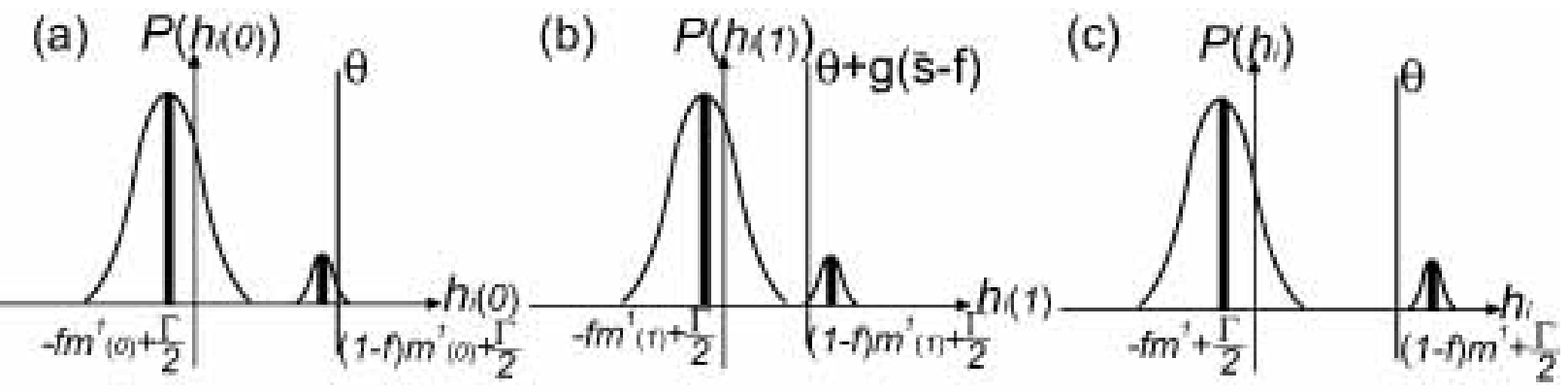}\\
\vspace{1.5cm}
{\Huge Figure 8}\\
\vspace{1cm}
\Large Matsumoto
\end{center}
\end{figure}

\clearpage
\begin{figure}[p]
\begin{center}
\includegraphics{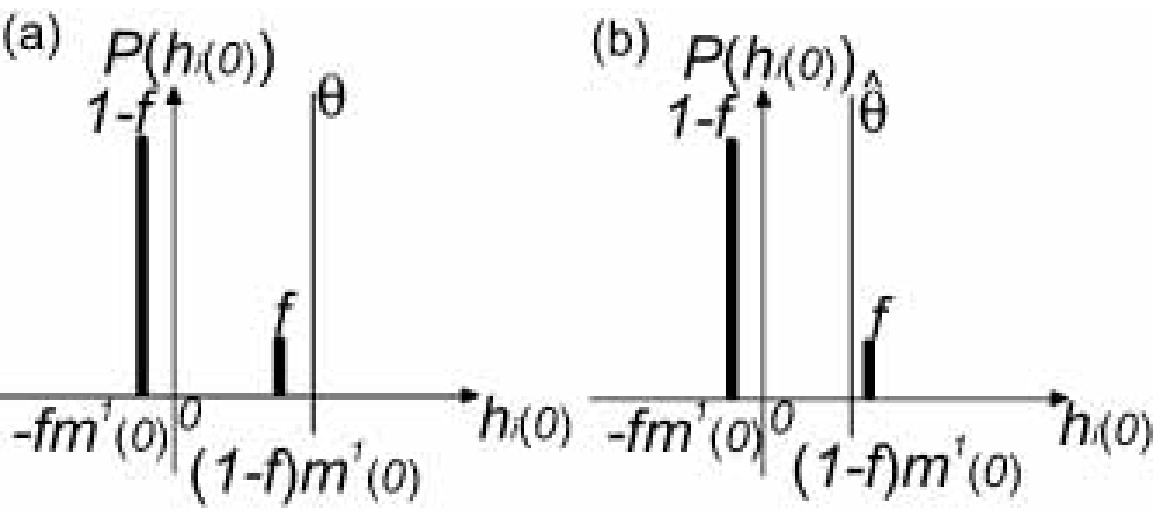}\\
\vspace{1.5cm}
{\Huge Figure 9}\\
\vspace{1cm}
\Large Matsumoto
\end{center}
\end{figure}

\clearpage
\begin{figure}[p]
\begin{center}
\includegraphics{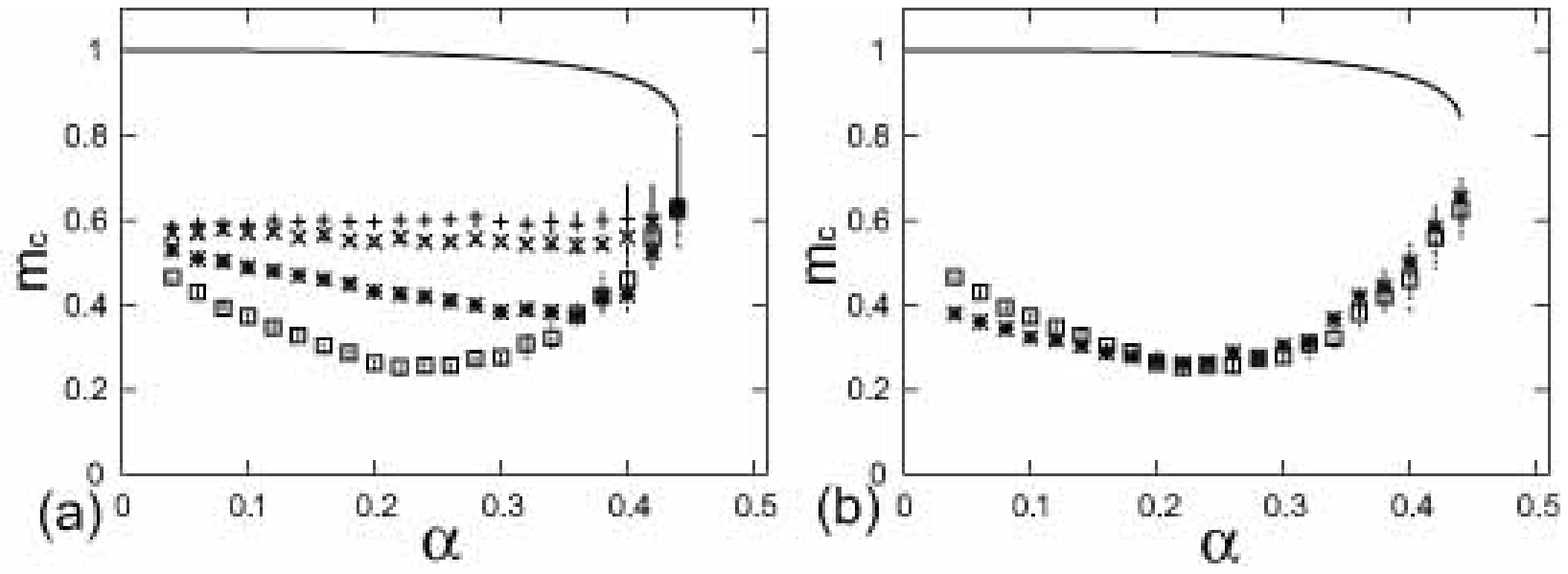}\\
\vspace{1.5cm}
{\Huge Figure 10}\\
\vspace{1cm}
\Large Matsumoto
\end{center}
\end{figure}

\end{document}